\documentclass[12pt,a4paper,final]{iopart}

\usepackage{graphicx,color}

\begin{document}

\title[Tunable single photon emission from dipolaritons]{Tunable single photon emission from dipolaritons}

\author[cor1]{O. Kyriienko$^{1,2}$, I. A. Shelykh$^{1,2}$, T. C. H. Liew$^{2}$}
\address{$^1$Science Institute, University of Iceland, Dunhagi-3, IS-107, Reykjavik, Iceland}
\address{$^2$Division of Physics and Applied Physics, Nanyang Technological University
637371, Singapore}
\ead{kyriienko@ukr.net}
\vspace{2pc}

\begin{abstract}
We study a system comprising of a double quantum well embedded in a micropillar optical cavity, where strong coupling between a direct exciton, indirect exciton, and cavity photon is achieved. We show that the resulting hybrid quasiparticles --- dipolaritons --- can induce strong photon correlations and lead to anti-bunched behaviour of the cavity output field. The origin of the observed single photon emission is attributed to unconventional photon blockade. Moreover, we find that the second-order equal time correlation function $g^{(2)}(0)$ can be tuned over a large range using an electric field applied to the structure, or changing the frequency of the pump. This allows for an on-the-flight control of cavity output properties, and is important for the future generation of tunable single photon emission sources.
\end{abstract}



\section{Introduction}
The generation of non-classical states of photons \cite{DavidovichRev,Shields2007} represents a hot research topic of modern quantum optics and is vital for future applications. In particular, single photon emitters (SPEs) are of key importance in quantum communication \cite{ScaraniRev}, quantum metrology \cite{LloydRev}, and quantum information technologies \cite{Knill2001,MilburnRev}, where strong anti-bunching of photons (and sub-Poissonian statistics) is required.

At the same time, a possibility to induce non-classical behaviour of photons ultimately relies on the effective photon-photon interactions, which bring nonlinearity into otherwise linear optical systems. Therefore, considering the typically small values of photon nonlinearity presently available, the search for an ideal system still persists \cite{CarusottoRev}. Moreover, an issue of tunability often appears as a compulsory demand for future SPEs, accounting for the strong dependence of emission properties on the system parameters.

Examples of currently existing and recently proposed single photon emitters contain a large number of various nonlinear systems, spanning from cavity quantum electrodynamic (cQED) systems with a single atom \cite{Imamoglu1997,Kuhn2002} or a quantum dot (QD) \cite{Reinhard2012,He2013}, photonic blockade by Rydberg atoms \cite{Gorshkov2011,Peyronel2012}, to four-level atomic systems \cite{Hartmann2007}, quantum optomechanical setups \cite{Nunnenkamp2011,Rabl2011}, confined cavity polaritons \cite{Verger2006}, and others \cite{Brunel1999,Kurtseifer2000,Kuzmich2000,Lounis2000}. Many of them depend on the strength of optical nonlinearity $U$, accounting for the fact that large nonlinearity makes the photon energy spectrum non-equidistant, with consequent suppression of two photon state occupation. This effect was coined as photon blockade \cite{Imamoglu1997}, given the similarity with electron blockade in quantum dot systems. However, an additional constraint is given by the cavity photon decay rate $\kappa$, demanding $U/\kappa \gg 1$ and posing the challenge of production of high finesse cavities or very large values of nonlinearity.

An alternative unconventional photon blockade effect was recently proposed \cite{Liew2010}. The origin of anti-bunching there lies in quantum interference effects \cite{Bamba2011}, and the condition $U/\kappa \gg 1$ is relaxed, while a small optical nonlinearity is still required for SPE operation. With an initial system of interest corresponding to tunnel-coupled polaritonic micropillars \cite{Liew2010,Bamba2011,Bamba2011APL,Flayac2013}, with significant experimental progress \cite{Vasconcellos2011} a number of studies reported similar effects in cQED systems with QDs \cite{Majumdar2012,Majumdar2012b}, coupled optomechanical cavities \cite{Komar2013,Xu2013,Savona2013}, doubly resonant microcavities \cite{Gerace2014} or passive nonlinear cavities \cite{Ferretti2012,Ferretti2013}. Other interesting proposals which also enable SPE for modest values of nonlinearity include weakly coupled optomechanical systems with enhanced parametric interaction \cite{Borkje2013,Lemonde2013} or critically coupled hybrid optomechanical systems \cite{Nori2013}.

Here, we show that an unconventional photon blockade can be engineered in a dipolariton system \cite{Cristofolini2012,Christmann2011}, where direct exciton and indirect exciton \cite{High2012} transitions of a double quantum well (QW) are strongly coupled to each other, and direct excitons are additionally strongly coupled to a cavity mode of a zero-dimensional micropillar [see sketch in Fig. 1(a)]. The emergent dipolariton quasiparticles were shown to be useful for the generation of THz radiation \cite{Kyriienko2013,Kristinsson2013,Kristinsson2014} and enhancement of the nonlinear interparticle interaction due to its dipole-dipole nature \cite{Cristofolini2012}. Assuming low-intensity coherent pumping of the cavity mode, we perform master equation calculations and observe anti-bunching of photons originating from coherent coupling between the modes. We define the optimal parameters for the single photon emission using an ansatz wave function in a truncated Fock space of the dipolariton system. Finally, we show that the second order equal time correlation function of photons can be easily tuned via an applied electric field or tuning the pumping frequency. This enables on-the-flight change of SPE properties necessary for future applications.

\section{Model}
We consider the system of a semiconductor micropillar~\cite{Bloch1997,Kaitouni2006,Bajoni2007} with a double quantum well embedded in the antinode of the optical resonator [Fig. 1(a)]. The quantum wells (QWs) are strongly coupled via electronic tunneling $J$, while the direct excitons from the left quantum well are additionally strongly coupled to the cavity mode (C) with Rabi frequency $\Omega$. The detuning between indirect exciton (IX) and direct exciton (DX) levels can be conveniently tuned using external electric field $F$.
\begin{figure}
\begin{center}
\includegraphics[width=1.0\linewidth]{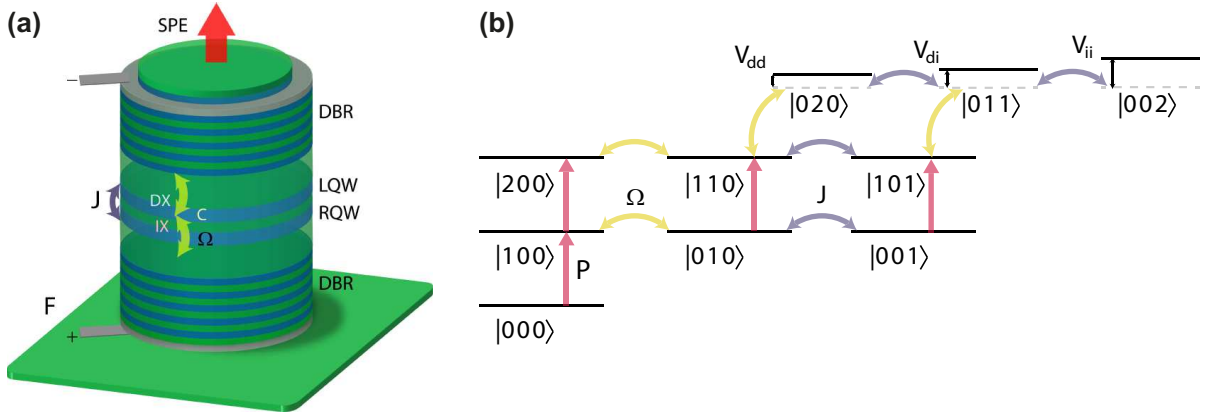}
\caption{(color online) (a) Sketch of the system showing a micropillar optical cavity formed by distributed Bragg reflectors (DBRs), with an embedded double quantum well (LQW and RQW). The direct exciton mode in the LQW is strongly coupled to the cavity field with Rabi frequency $\Omega$. Also, the applied electric field $F$ is tuned such that the direct exciton and indirect exciton modes are strongly coupled via tunneling with rate $J$. (b) The sketch of the joined cavity photon--direct exciton--indirect exciton Fock space showing the relevant transition processes between states, leading to the unconventional photon blockade. Here yellow ($\Omega$) and blue ($J$) arrows correspond to linear couplings, red arrows ($P$) denote weak optical pump, and $V_{dd,di,ii}$ are energy shifts due to nonlinearity.}
\end{center}
\label{fig:sketch}
\end{figure}

The generic Hamiltonian for the dipolariton system reads $\hat{\cal{H}}=\hat{\cal{H}}_{0} + \hat{\cal{H}}_{int}$, and consists of coherent linear and nonlinear contributions. The linear part of the Hamiltonian reads:
\begin{eqnarray}
\label{H_0}
\hat{\cal{H}}_{0}=\hbar \omega_{C}\hat{a}^{\dagger}\hat{a}+\hbar \omega_{DX}\hat{b}^{\dagger}\hat{b}+\hbar \omega_{IX}\hat{c}^{\dagger}\hat{c}+\frac{\hbar \Omega}{2}(\hat{a}^{\dagger}\hat{b}+\hat{b}^{\dagger}\hat{a})\nonumber\\
\hspace{20mm}-\frac{\hbar J}{2}(\hat{b}^{\dagger}\hat{c}+\hat{c}^{\dagger}\hat{b})+Pe^{-i\omega_p t}\hat{a}^{\dagger}+P^{*}e^{i\omega_p t}\hat{a},
\end{eqnarray}
where $\hat{a}^{\dagger},~\hat{a}$, $\hat{b}^{\dagger},~\hat{b}$ and $\hat{c}^{\dagger},~\hat{c}$ are creation and annihilation operators for cavity photons, direct excitons and indirect excitons, respectively. Here $\hbar \omega_{C}$, $\hbar \omega_{DX}$ and $\hbar \omega_{IX}$ denote the cavity mode, direct exciton and indirect exciton energies, and the first three terms of the Hamiltonian describe the energy of the bare modes. The fourth and fifth terms correspond to the strong coupling between modes, where $\hbar \Omega $ (Rabi splitting) denotes the coupling constant between photons and direct excitons, and the tunnelling rate corresponding to DX--IX coupling is $\hbar J$. The last two terms in the Hamiltonian correspond to the optical pumping of the cavity mode with rate $P$ and frequency $\omega_p$. The relevant processes given by the coherent linear couplings and nonlinear shifts are schematically depicted in Fig. 1(b).

It is convenient to perform the unitary transformation $\hat{\mathcal{H}}_0' = \hat{U}^\dagger ( \hat{\mathcal{H}}_0 -i \hbar \partial_t )\hat{U}$, which redefines the overall energy scale and removes the time dependence of the pump term. The unitary operator reads $\hat{U} = \exp[-i\omega_p t (\hat{a}^\dagger\hat{a} + \hat{b}^\dagger \hat{b} + \hat{c}^\dagger \hat{c})]$, and it is useful to introduce the relevant detuning parameters $\Delta \equiv \omega_p - \omega_C$, $\delta_{\Omega} \equiv \omega_{C}-\omega_{DX}$, and $\delta_{J} \equiv \omega_{IX}-\omega_{DX}$. Implementing rotation, the Hamiltonian reads
\begin{eqnarray}
\label{H_0_RWA}
\hat{\cal{H}}_{0}' = -\hbar\Delta\hat{a}^{\dagger}\hat{a}-\hbar(\delta_{\Omega}+\Delta)\hat{b}^{\dagger}\hat{b}+\hbar(\delta_J -\delta_{\Omega}-\Delta)\hat{c}^{\dagger}\hat{c}+\frac{\hbar \Omega}{2}(\hat{a}^{\dagger}\hat{b}+\hat{b}^{\dagger}\hat{a})\nonumber\\
\hspace{20mm}-\frac{\hbar J}{2}(\hat{b}^{\dagger}\hat{c}+\hat{c}^{\dagger}\hat{b})+P\hat{a}^{\dagger}+P^{*}\hat{a}.
\end{eqnarray}
Note that the photon-exciton detuning $\delta_{\Omega}$ can be controlled during the growth stage by preparing micropillar cavities of different length, and the IX-DX detuning $\delta_{J}(F) = \delta^{(0)}_{J}(1- F/F_0)$ can be easily tuned by an applied electric field $F$, where $\delta^{(0)}_{J}= \omega_{IX}(0)-\omega_{DX}(0)$ is an exciton detuning at zero applied field and $F_0$ is electric fiend corresponding to the crossing of the modes \cite{Christmann2011}.

The nonlinear interactions between excitons are contained in the $\hat{\cal{H}}_{int}$ Hamiltonian, given by
\begin{equation}
\label{H_int}
\hat{\cal{H}}_{int}= V_{dd} \hat{b}^{\dagger}\hat{b}^{\dagger}\hat{b}\hat{b} + V_{ii} \hat{c}^{\dagger}\hat{c}^{\dagger}\hat{c}\hat{c} + V_{di} \hat{b}^{\dagger}\hat{c}^{\dagger}\hat{b}\hat{c},
\end{equation}
where $V_{dd}$, $V_{ii}$ and $V_{id}$ denote direct exciton, indirect exciton, and direct-indirect exciton interaction strengths.

In general, the Hamiltonian of the real dipolariton system involves both coherent and decoherent parts. The latter can be treated within the master equation for the density matrix $\rho$ written in the joint C-DX-IX Fock space, being $| n_{a} n_{b} n_{c} \rangle = | n_a \rangle \otimes | n_b \rangle \otimes | n_c \rangle$. The master equation reads
\begin{eqnarray}
i\hbar \frac{\partial \rho}{\partial t}=[\hat{\cal{H}},\rho]+i\frac{\kappa}{2}\mathcal{D}[\hat{a}] +i\frac{\gamma_{dx}}{2} \mathcal{D}[\hat{b}] +i\frac{\gamma_{ix}}{2} \mathcal{D}[\hat{c}] +i\frac{\Gamma^{(dec)}_{X}}{2} \mathcal{D}[\hat{c}^\dagger \hat{c}],
\label{ME}
\end{eqnarray}
where the superoperator $\mathcal{D}[\hat{\mathcal{O}}] = 2\hat{\mathcal{O}}\rho\hat{\mathcal{O}}^\dagger- \{\hat{\mathcal{O}}^\dagger \hat{\mathcal{O}},\rho\}$ corresponds to a dissipator written in the Lindbladian form. Here $\kappa$, $\gamma_{dx}$ and $\gamma_{ix}$ correspond to dissipation rates of the cavity photon, direct exciton, and indirect exciton mode, respectively. The last term in Eq. (\ref{ME}) describes pure dephasing in the system related to exciton scattering processes, characterised by the dephasing rate $\Gamma^{(dec)}_{X}$. While this term does not affect the occupation of the modes, it influences the temporal dynamics of the system, and should be accounted for in delay-dependent calculations.

Solving the dynamical equation to calculate the evolution of the density matrix $\rho(t)$, one can easily find the mean values of number operators $\langle \hat{N}(t) \rangle = \langle \hat{a}^{\dagger} \hat{a} \rangle = \mathrm{Tr}\{ \hat{a}^{\dagger} \hat{a} \rho(t) \}$, and the second order correlation function of photons, $g^{(2)}(\tau)= \langle \hat{a}^\dagger(t) \hat{a}^\dagger(t+\tau) \hat{a}(t+\tau) \hat{a}(t) \rangle / \langle \hat{N}(t) \rangle \langle \hat{N}(t+\tau) \rangle$.

\section{Optimisation of parameters}
While master equation simulations provide full information about the system behaviour, it is instructive to use a simplified approach based on a trial wave function, written in the form \cite{Bamba2011,Komar2013}:
\begin{eqnarray}
|\Psi \rangle = A_{000} |000\rangle + A_{100} |100\rangle + A_{010} |010\rangle + A_{001} |001\rangle + A_{200} |200\rangle  \\ \nonumber \hspace{10mm} + A_{110} |110\rangle + A_{101} |101\rangle + A_{011} |011\rangle + A_{020} |020\rangle + A_{002} |002\rangle + ...,
\end{eqnarray}
where $A_{ijk}$ represents the amplitude of the state, and we consider only the lowest Fock states, assuming weak optical pumping. We can consider evolution of the system for the effective non-hermitian Hamiltonian $\hat{\mathcal{H}}_{eff}= \hat{\mathcal{H}}_{0}' + \hat{\mathcal{H}}_{int} - i\kappa \hat{a}^\dagger \hat{a}/2$, where the last term accounts for the decay of the cavity mode. Plugging the trial solution into the Schr\"{o}dinger equation $\hat{\mathcal{H}}_{eff}|\Psi\rangle = i\hbar d |\Psi \rangle /dt = 0$, and assuming that the system is initially prepared in the $|000\rangle$ state, we can find steady state values of the amplitudes. Moreover, accounting for the weak pumping conditions, amplitudes can be found in an iterative manner, meaning that $A_{000}\gg A_{100},A_{010},A_{001}\gg A_{200},A_{110},A_{101},A_{011},A_{020},A_{002}$. The system of steady-state equations for amplitudes can be solved symbolically. However, the resulting expressions are too large to be presented here, and can only be analysed numerically.
\begin{figure}
\begin{center}
\includegraphics[width=0.8\linewidth]{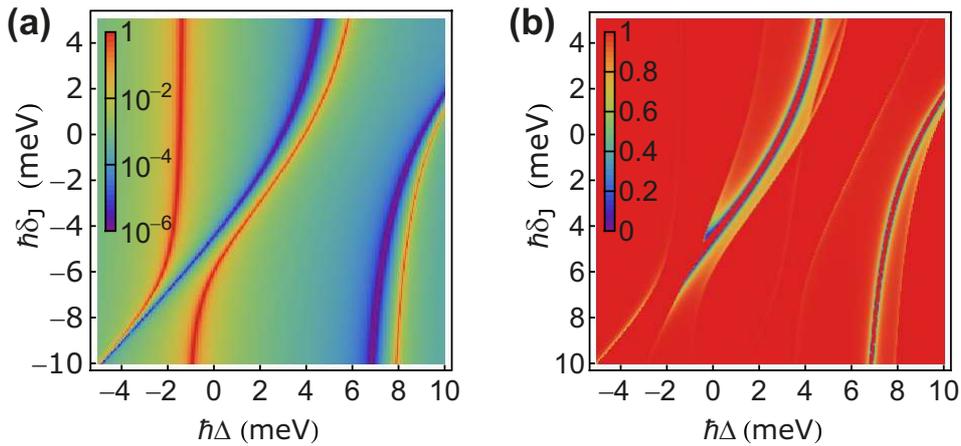}
\caption{(color online) (a) and (b): Density plots of the mean photon occupation number and the second-order correlation function calculated for different values of pump detuning $\Delta$ and exciton detuning $\delta_J(F)$, which can be controlled by the applied electric field.}
\end{center}
\label{fig:density_plots}
\end{figure}

The optimisation procedure can be realised as follows. From the six equations for the two-particle state amplitudes we are mostly interested in the two-photon state one, $A_{200}$. It is given by a function of system parameters $A_{200}(\Omega,J,\kappa,V_{dd},V_{ii},V_{id},\delta_{\Omega};\Delta,\delta_{J})$, where the Rabi frequency $\Omega$, photon decay rate $\kappa$, exciton interaction constants, tunneling coupling $J$ and cavity detuning $\delta_{\Omega}$ are intrinsic parameters of the sample which can be changed during the micropillar growth process only. In particular, the tunneling coupling $J$ is adjusted choosing the separation between quantum wells, and the cavity photon detuning $\delta_{\Omega}$ is controlled by the cavity length $L_{cav}$ (or preparing an array of micropillars with different $L_{cav}$). At the same time, the detuning of the pumping laser $\Delta$ can be changed in experiment using a source with adjustment of frequency and the IX-DX detuning $\delta_{J}(F)$ is electric-field dependent, leading to a high degree of control.
To achieve an optimal anti-bunching of photons one needs to find the parameters which yield nulling of both real and imaginary parts of the two-photon amplitude $A_{200}(\Omega,J,\kappa,V_{dd},V_{ii},V_{id},\delta_{\Omega};\Delta,\delta_{J})$ due to interference of different excitation paths [Fig. 1(b)]. Thus, we naturally choose $\Delta$ and $\delta_J$ as optimisation parameters, while fixing the others. Solving the system of two equations for optimal detuning $\Delta^{(opt)}$ and $\delta_{J}^{(opt)}$, we can find that several branches of optimal solutions appear in a certain range of parameters.

\section{Results and discussion}

We consider a micropillar of $2~\mu$m diameter with an embedded In$_{0.1}$Ga$_{0.9}$As/GaAs/In$_{0.08}$Ga$_{0.92}$As double quantum well, where each QW has 10 nm width and the spacer width is 4 nm \cite{Cristofolini2012}. The corresponding tunneling coupling and Rabi frequency of the exciton-photon coupling are $\hbar J =\hbar \Omega = 6$ meV. The decay rate of the cavity mode is $\kappa=0.2$ meV and we set the pumping to $P=0.1$ meV. The non-radiative exciton decay rates $\gamma_{dx}$ and $\gamma_{ix}$ are typically three orders of magnitude smaller that $\kappa$, and therefore we disregard them in master equation calculations. The exciton interaction constants for the given system can be estimated as $V_{dd}=0.0015$ meV \cite{Tassone1999}, $V_{ii}=0.025$ meV \cite{Kyriienko2012}, and $V_{di}=0.017$ meV \cite{Kristinsson2013}. The photonic detuning $\hbar \delta_{\Omega}$ can be chosen in a wide range, while we note that the described photon blockade processes favour a negative value. In further calculations we fix $\hbar \delta_{\Omega}=-6$ meV.
\begin{figure}
\begin{center}
\includegraphics[width=0.8\linewidth]{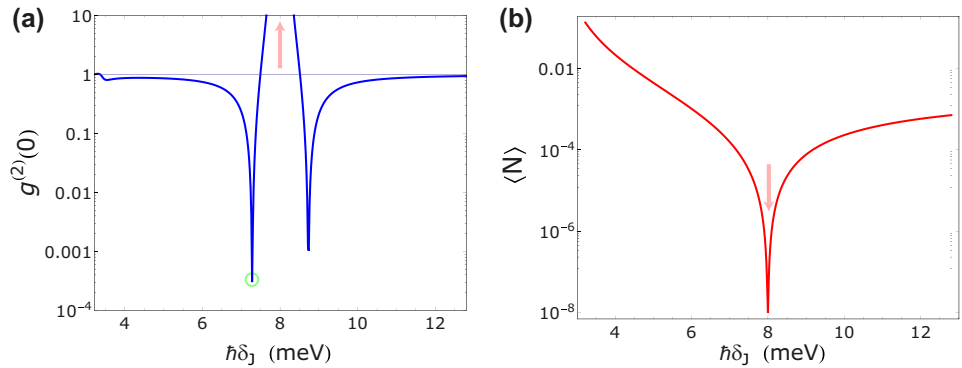}
\caption{(color online) (a) Second order coherence function for zero delay $g^{(2)}(0)$ shown as a function IX-DX exciton detuning $\delta_J(F)$ for a fixed pump frequency $\hbar \Delta = 5$ meV, and pump intensity $P = 0.8$ meV. Green circle points the region of strong antibunching. (b) Mean photon value $\langle N \rangle$ plotted for the same range of detuning. The red arrow shows the point where the mean photon number greatly reduces.}
\end{center}
\label{fig:g2_dJ}
\end{figure}

Next we perform master equation calculations using the truncated Fock space, where up to five particle states are retained for each mode. The details of the calculation algorithm can be found in Refs. \cite{Liew2010,Liew2012}. Finding the steady state density matrix of the system, we can calculate the photon occupation number and second-order correlation function for various values of pump and excitonic detunings. The results are shown in Fig. 2(a,b). We note that there are several dips of $g^{(2)}(0)$ associated to anti-bunching, which can be found in a broad range of parameters. Finding these minima, one can also extract the optimal parameters which should be followed in order to have single photon emission.
\begin{figure}
\begin{center}
\includegraphics[width=0.8\linewidth]{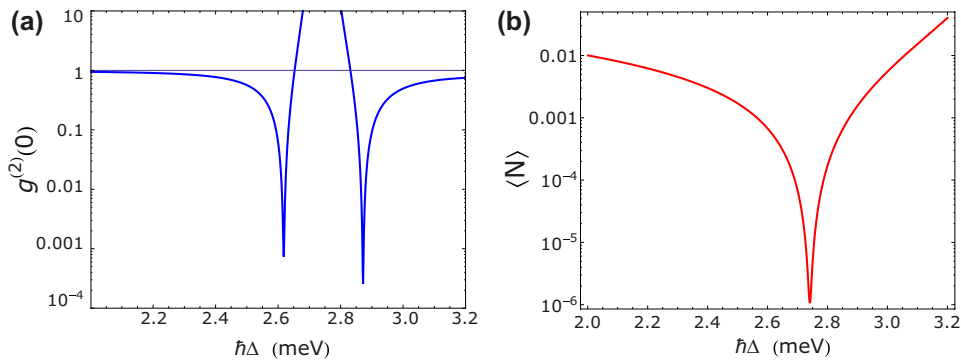}
\caption{(color online) (a) and (b): Second order coherence function $g^{(2)}(0)$ and mean photon number $\langle N \rangle$ are plotted as a function pump frequency $\hbar \Delta$ for a fixed IX-DX exciton detuning $\delta_J(F)=-0.5$ meV and $P = 0.8$ meV.}
\end{center}
\label{fig:g2_D}
\end{figure}

Using the optimisation procedure outlined before, we find the optimal pump and exciton detunings $\hbar \Delta^{(opt)}=4.4$ meV and $\hbar \delta_J^{(opt)}=3.57$ meV, with the minimal value of second order coherence function $g^{(2)}_{min}(0) \approx 1\times 10^{-4}$. Thus, we confirm that the dipolariton system indeed can serve as a setup where strong effective optical nonlinearity can be attained. Finally, it is instructive to fix the pumping energy, e.g., to $\hbar \Delta = 5$ meV and plot $g^{(2)}(0)$ as a function of $\delta_J$. The result is shown in Fig. 3, where both equal time second order coherence functions and mean photon numbers are plotted. We observe two slightly asymmetric minima, in the form of an inverted ``batman'' lineshape. The unconventional blockade window $\xi(\delta_J)$, defined as the detuning region at which $g^{(2)}(0)$ drops below the $10\%$ level, can be estimated as $\xi(\delta_J)=0.26$ meV. Here an important improvement over previously considered systems is the possibility of on-the-flight control of the emission coherence properties, where IX-DX detuning is conveniently tuned by an electric field $F$ applied to the structure. Additionally, we find that there are bunching regions accompanied with a strong reduction of the photon mean number (marked by red arrows in Fig. 3).

Also, we can perform a similar procedure, fixing the exciton detuning $\hbar \delta_J = -0.5$ meV. The corresponding cross-section of the density plot is shown in Fig. \ref{fig:g2_D}, showing that the statistics of the optical field emitted from the dipolariton system can be also controlled by the frequency of the laser pump.

Another quantity which characterises the single photon emitter is the second-order correlation function calculated for finite delay, $g^{(2)}(\tau)$. In particular, the positive derivative of this function with respect to variable $\tau$ represents a nonclassical behavior of the emitter. Additionally, it is important for the experimental observation of the antibunching, where precise measurement of the equal time correlation function may be challenging.
\begin{figure}
\begin{center}
\includegraphics[width=0.7\linewidth]{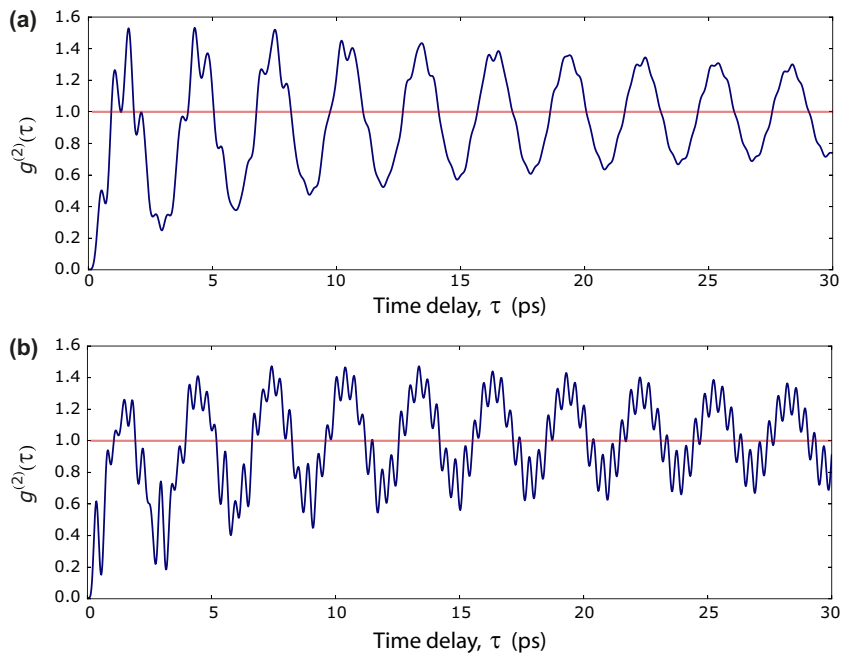}
\caption{(color online) (a) and (b): Second order correlation function at finite delay $\tau$ calculated for different values of the pure dephasing of the excitons, being $\Gamma^{(dec)}_{X} = 0$ (a) and $\Gamma^{(dec)}_{X} = 0.3~\mu$eV (b).}
\end{center}
\label{fig:g2tau}
\end{figure}

Using the master equation and quantum regression theorem we calculate the function $g^{(2)}(\tau)= \langle \hat{a}^\dagger(0) \hat{a}^\dagger(\tau) \hat{a}(\tau) \hat{a}(0) \rangle / \langle \hat{N}(0) \rangle \langle \hat{N}(\tau) \rangle$ starting from the steady state density matrix defined as $\rho(0)$ \cite{Johansson2013}. The result is shown in Fig. 5, where figures (a) and (b) were plotted for different values of excitonic dephasing $\Gamma^{(dec)}_{X}$. Similarly to the case of two coupled polaritonic boxes, $g^{(2)}(\tau)$ represents an oscillatory function with a period determined by the coupling parameters $J$ and $\Omega$. Assuming the excitonic pure dephasing to be absent, we find that $g^{(2)}(\tau)$ experiences damped short period oscillations of $2 \pi/J$ order, with an additional long period harmonic behavior ($T\approx$ 4 ps), shown in Fig. 5(a). However, accounting for the associated dephasing of excitons coming from the Coulomb scattering, the calculated second order coherence function exhibits more pronounced short-scale oscillations [Fig. 5(b)]. This can be related to the fact that while IX-DX Rabi flopping is disrupted, the behaviour is mostly governed by the coherent C-DX oscillations of the period $2\pi/\Omega$.

\section{Conclusion}
We have considered a system of dipolaritons formed in a micropillar optical cavity. Accounting for the linear couplings between cavity photon, direct exciton, and indirect exciton modes, and intrinsic nonlinearity for the excitons, we find associated unconventional photon blockade, leading to anti-bunching of the order $g^{(2)}(0) \sim 10^{-4}$. Given the large parameter space of the system and its high sensitivity, we perform an optimisation procedure, where optimal pump detuning $\Delta^{(opt)}$ and IX-DX detuning $\delta_J^{(opt)}$ can be easily adjusted to generate unconventional photon blockade. The current study thus facilitates the way towards tunable single photon sources relevant for various applications.\\

The work was supported by Tier 1 project ``Polaritons for novel device applications,'' FP7 IRSES projects POLAPHEN and POLATER, and FP7 ITN NOTEDEV network. O.~K. acknowledges the support from the Eimskip fund.

\end{document}